%% file: hmendez2_cleo_main.tex
\documentclass[
   ,final            
  ,numberedheadings 
,sort&compress   
  ]
  {aipproc}

\layoutstyle{6x9}

\usepackage{xspace}

\include{hmendez2_cleo_def}

\begin{document}

\title{R\lowercase{ECENT} \cleoc R\lowercase{ESULTS}}


\author{H. Mendez\\(for the CLEO Collaboration)}{
  address={Department of Physics, University of Puerto Rico,
           Mayaguez, PR 00681\\
	   E-mail: mendez@charma.uprm.edu}
}



\begin{abstract}
Recent \cleoc results on open and closed charm physics
at center-of-mass energy of
3773~MeV (\psidd resonance),
4170~MeV and
3686~MeV (\psip peak)
are reviewed.
Measurements of
absolute hadronic branching ratios of \ddo, \ddp and \ddsP mesons
as well as
charmonium spectroscopy are discussed. 
An outlook and future prospects for the experiment
at CESR is also presented.


\end{abstract}

\maketitle


\section{I\lowercase{ntroduction}}\label{Intro}
Recent advances on precise numerical calculations
of non-perturbative QCD,
formulated on a space-time lattice (LQCD)~\cite{Aubin:2004ej}
as well as the development of
heavy quark effective theory~\cite{Mannel:1996cn}
have produced a wide variety of non-perturbative results with
high
accuracies 
for D meson decay constants and form factors.
High precision charm experiments~\cite{yellowbook}
provide crucial data to validate
forthcoming LQCD calculations at the few percent level
and help to guide the development of QCD calculations techniques
for a full understanding of non-perturbative QCD effects.
These calculations, useful
to improve the accuracy of the 
Cabibbo-Kobayashi-Maskawa (CKM) unitary matrix~\cite{Kobayashi:1973fv},
make possible alternate methods to improve the precision
of the Standard Model predictions of charm behavior.
Thus, experiments on charm decays are an excellent
laboratory for testing
the existing theory on heavy flavor physics.

\cleoc is a dedicated program of charm physics
at the Cornell Electron Storage Ring (CESR)
\epem collider
located at the Laboratory for Elementary Particles Physics (LEPP)
at Cornell University at Ithaca, New York.
The experiment~\cite{yellowbook}
is designed to make very high precision measurements of charmed mesons
(\ddo, \ddp and \ddsP)\footnote{Charge conjugate particles are implicitly
assumed in this paper, unless otherwise noted.} 
and to test quantum chromodynamics (QCD),
which included a complete set of measurements 
for hadronic, leptonic and semi-leptonic charm decays,
a detailed studies on 
the lowest and highest mass charmonium states
and
search for evidence of new physics beyond the Standard Model
by  searching for
rare D and \ttau decays, \ddmix mixing, and CP violating decays.
\cleoc precision measurements on charmed meson decays to
leptonic and semi-leptonic
final states are crucial test of the LQCD techniques to compute important
heavy quark processes.
Hadronic decays play an important role for B physics branching fractions
normalization
as well as
in the study of final state which included strong interactions.

The experimental conditions in the charm system at threshold are optimal.
Charm events produced at threshold are extremely clean and
pure \ddmix events,
signal/background ratio is high and neutrino reconstruction is clean.
In this report, we review some of our recent results on
absolute hadronic branching ratio on \ddo, \ddp and \ddsP
(section~\ref{dd} and~\ref{ds}),
charmonium (section~\ref{cc})
and a brief description of the detector (section~\ref{Spec}). 

\section{E\lowercase{xperimental} S\lowercase{etup and} D\lowercase{ata}
               S\lowercase{ample}}\label{Spec}
In order to achieve this broad physics program,
both the CESR accelerator 
and the \cleoiii detector were upgraded.
The CESR \epem collider was converted to operate 
from $\sim$10~GeV 
to a lower
center-of-mass energies ($\sim$3.6-4.3~GeV) by the
addition of 18 meters of
superconducting wiggler magnets to enhance the 
transverse cooling of the beam.
In the \cleoiii detector,
the silicon vertex detector was replaced with a
6-layer vertex drift chamber
and
the solenoid field was reduced from 1.5 to 1.0~T,
which was achieved by a simple reduction of the magnet current.

\begin{figure}[!h]
  \centering
  \includegraphics[height=4.0in,angle=-90]
          {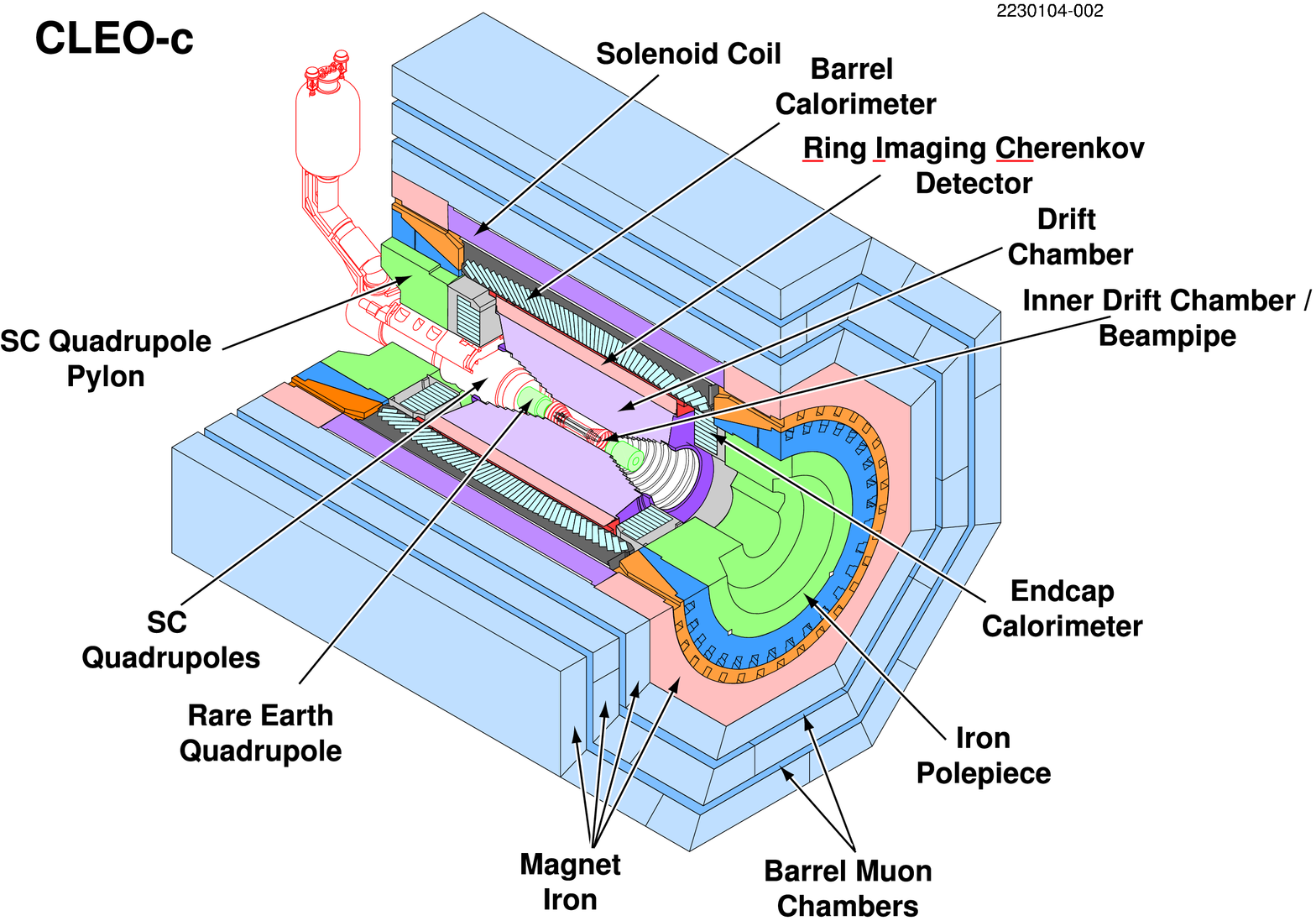}
  \caption{\cleoc Detector\label{CLEOcSpec}}
\end{figure}

The \cleoc spectrometer
(described extensively elsewhere~\cite{CLEOcDetec1,CLEOcDetec2,Kubota:1991ww,
Artuso:2005dc,Artuso:2002ya,Peterson:2002sk}),
shown in Fig.~\ref{CLEOcSpec}
is composed of the following basic sections:
\begin{enumerate}
\item a tracking system composed of the inner vertex drift chamber,
      a 47-layer central drift chamber~\cite{Peterson:2002sk}
      and a
      superconducting solenoid
      running at 1.0~T magnetic field, oriented along the beam axis.
      The system covers 93\% of 4$\pi$ and
      has a momentum resolution of 
      $\sigma_p/P~=~0.6$\% at 1~GeV/c,    
\item a cylindrical Ring Imaging Cerenkov (RICH)~\cite{Artuso:2005dc} detector,
      surrounding the central drift chamber,
      for charged particle identification.
      The RICH
      cover 80\% of 4$\pi$ and
      its efficiency to identified kaon is
      higher than 90\%
      with 0.2\% pion fake at P = 0.9~GeV/c,
\item an Electromagnetic Calorimeter~\cite{Kubota:1991ww}
      for neutral particle reconstruction
      covering 93\% of 4$\pi$.
      The detector has an
      energy resolution of $\sigma_E/E~=~2.2$\% at 1~GeV
      and 5\% at 100~MeV,
\item and a muon chambers
      covering 85\% of 4$\pi$ for P~$>$~1~GeV/c.
\end{enumerate}

The \cleoc detector is fully operational and presently taking data
at the \psidd center-of-mass energy.

A detailed GEANT~\cite{geant} based Monte Carlo
detector modeling was used 
to simulate the performance of the detector.
Physics events were 
generated by the
event generator EvtGen~\cite{Evtgen}
and final state radiation (FSR)
modeled by PHOTOS~\cite{Photos}.
Monte Carlo events were treated in the same manner as data. 

Charged particle candidates were required to be well measured
and to satisfy criteria based on the track fit quality.
They must also be
consistent with coming from the interaction point,
except for the charged pions coming from \kshor decay.
Particle identification to separate charged $\pi$ from K
was accomplished
by combining the ionization energy loss (dE/dx),
measured by the
drift chamber, with the RICH detector information.

In addition, we applied mass bound cut and kinematic fit to select
\kshor, \pizero, \etz and \etp.
We selected \kshor candidates from oppositely-charged tracks consistent
with pions and constrained to the \kshor vertex (\kshordec).
The \pizero candidates are form by a photon pair (\piodecay) kinematically
fitted to the nominal \pizero mass~\cite{PDG2006}.
Both of the invariant masses,
for the \kshor(\pipi) and for the \pizero(\phot\phot) candidates
are required  to be within 3 standard deviations ($\sigma$) of their 
known mass~\cite{PDG2006}.
For \kshor, we required  $\sigma \approx 6.3~MeV/c^2$ and
for \pizero, we required $\sigma \approx 5-7~MeV/c^2$
depending of the photon location in the calorimeter.
We form \etz candidates from photon pairs kinematically fitted to the
known \etz mass~\cite{PDG2006} and
\etp candidates are detected via their decays mode \etpdec.

The experiment has already collected
281~pb$^{-1}$ of data at center-of-mass energy
\comeq~3.77~GeV to study \psidd decays,
approximately 300~pb$^{-1}$ at
\comeq~4.17~GeV to study \dds mesons  and
approximately 50~pb$^{-1}$ at
\comeq~3.68~GeV to study  \psip  decays and charmonium spectrum,
including \jpsi, \chicjn and the properties of the
\cleoc observed \hc state.
A small fraction (2.74~pb$^{-1}$) of the \psip integrated luminosity 
was taken with the \cleoiii detector while its conversion to \cleoc 
took place.
In addition, the experiment has recorded a significant amount of
continuum data to study its contribution to these resonances.


\section{D P\lowercase{hysics at} \psidd}\label{dd}
The 3773~MeV dataset,
taken at the peak of the \psidd resonance,
provides a very clean environment for studying with high precision
a great variety of D meson decays.
Many results based on an initial sample of 56~pb$^{-1}$
have already been published.
These results, among others, included
absolute branching fraction measurements for D decays into 
leptonic~\cite{Bonvicini:2004gv},
semi-leptonic~\cite{Huang:2005iv,Coan:2005iu},
and hadronic~\cite{He:2005bs} final states.
In this contribution, we report preliminary
hadronic branching fraction measurements based on the total \psidd sample
of 281~pb$^{-1}$
available at the present.
This dataset include approximately
1 million of \epemtodz and 0.8 million of \epemtodp events.

The \psidd produced in the \epem annihilation
decays to pairs of D mesons only, either \ddo$\overline{\ddo}$ or
\ddp\ddm.
At this energy, pure \ddmix final states are produced,
no additional hadron accompanied the D mesons,
therefore, the reconstructed D energy is safely replaced 
by the beam energy (E$_D~\equiv$~E$_{Beam}$) and
as a consequence,
the reconstructed
D invariant mass
is a beam constrained quantity
(m$_{bc}~=~\sqrt{E_{Beam}^2-\vec{P_D}^2}$~),
which has better resolution than the
invariant mass calculated using directly the energy of the D.
For equal mass particles of mass M as this case, 
the kinematic variables m$_{bc}$ peak at M 
and the energy difference $\Delta$E~$\equiv$~E$_D$~-~E$_{Beam}$
peak at zero.  
Both variables
were used
in the analysis of the \psidd data.
$\Delta$E around zero 
was required for the D candidates.

\cleoc uses a double and single tagging technique, pioneered by the 
MARK III collaboration~\cite{Baltrusaitis:1985iw,Adler:1987as}, 
to measure absolute branching fractions.
It relies on fully reconstructed \ddmix decays,
in which both D are reconstructed (double tags)
in the event.
Single tag mode reconstructs at least one D per event.
A full description of the tagging method used by the experiment
as well as the hadronic analysis based on 56~pb$^{-1}$ 
can be found in~\cite{He:2005bs}.
The tagging technique obviated the need for knowledge of the
luminosity or the \epemtodd production cross section.
The absolute branching fractions obtained using double tag~\cite{Cassel:HQL06}
are listed in Table~\ref{tab:DBranching}.
\begin{table}[!h]
\begin{tabular}{lr}
\hline
  \tablehead{1}{c}{b}{\ddo Mode}                  
  & \tablehead{1}{r}{b}{Branching Fraction (\%)}  
\\
\hline
\ddokpi     &  3.87 $\pm$ 0.04 $\pm$ 0.08 \\   
\ddokpipiz  & 14.6~ $\pm$ 0.1~ $\pm$ 0.4~ \\   
\ddokpipipi &  8.3~ $\pm$ 0.1~ $\pm$ 0.3~ \\   
\hline
  \tablehead{1}{c}{b}{\ddp Mode}
\\
\hline
\ddpkpipi    & 9.2~ $\pm$ 0.1~ $\pm$ 0.2~ \\
\ddpkpipipiz & 6.0~ $\pm$ 0.1~ $\pm$ 0.2~ \\
\ddpkspi     & 1.55 $\pm$ 0.02 $\pm$ 0.05 \\
\ddpkspipiz  & 7.2~ $\pm$ 0.1~ $\pm$ 0.3~ \\
\ddpkspipipi & 3.13 $\pm$ 0.05 $\pm$ 0.14 \\
\ddpkkpi     & 0.93 $\pm$ 0.02 $\pm$ 0.03 \\
\hline
  \tablehead{1}{c}{b}{\ddsP Mode}
\\
\hline
\ddsPksk      &  1.50 $\pm$ 0.09 $\pm$ 0.05 \\
\ddsPkkpi     &  5.57 $\pm$ 0.30 $\pm$ 0.19 \\
\ddsPkkpipiz  &  5.62 $\pm$ 0.33 $\pm$ 0.51 \\
\ddsPpipipi   &  1.12 $\pm$ 0.08 $\pm$ 0.05 \\
\ddsPpietz    &  1.47 $\pm$ 0.12 $\pm$ 0.14 \\
\ddsPpietp    &  4.02 $\pm$ 0.27 $\pm$ 0.30 \\
\hline
\end{tabular}
\caption{Preliminary absolute branching fractions for
         \ddo and \ddp based on 281~pb$^{-1}$
	 and
	 \ddsP based on the 195~pb$^{-1}$ sample.
	 The first error is statistical and the second is systematics.}
\label{tab:DBranching}
\end{table}


These results are dominated by the systematic errors
and included the FSR correction, which has been included in the 
Monte Carlo.
Without this effect, the branching fractions decreased in average
by 2\%.

\begin{figure}    
  \centering
  \includegraphics[height=2.18in,width=2.5in]{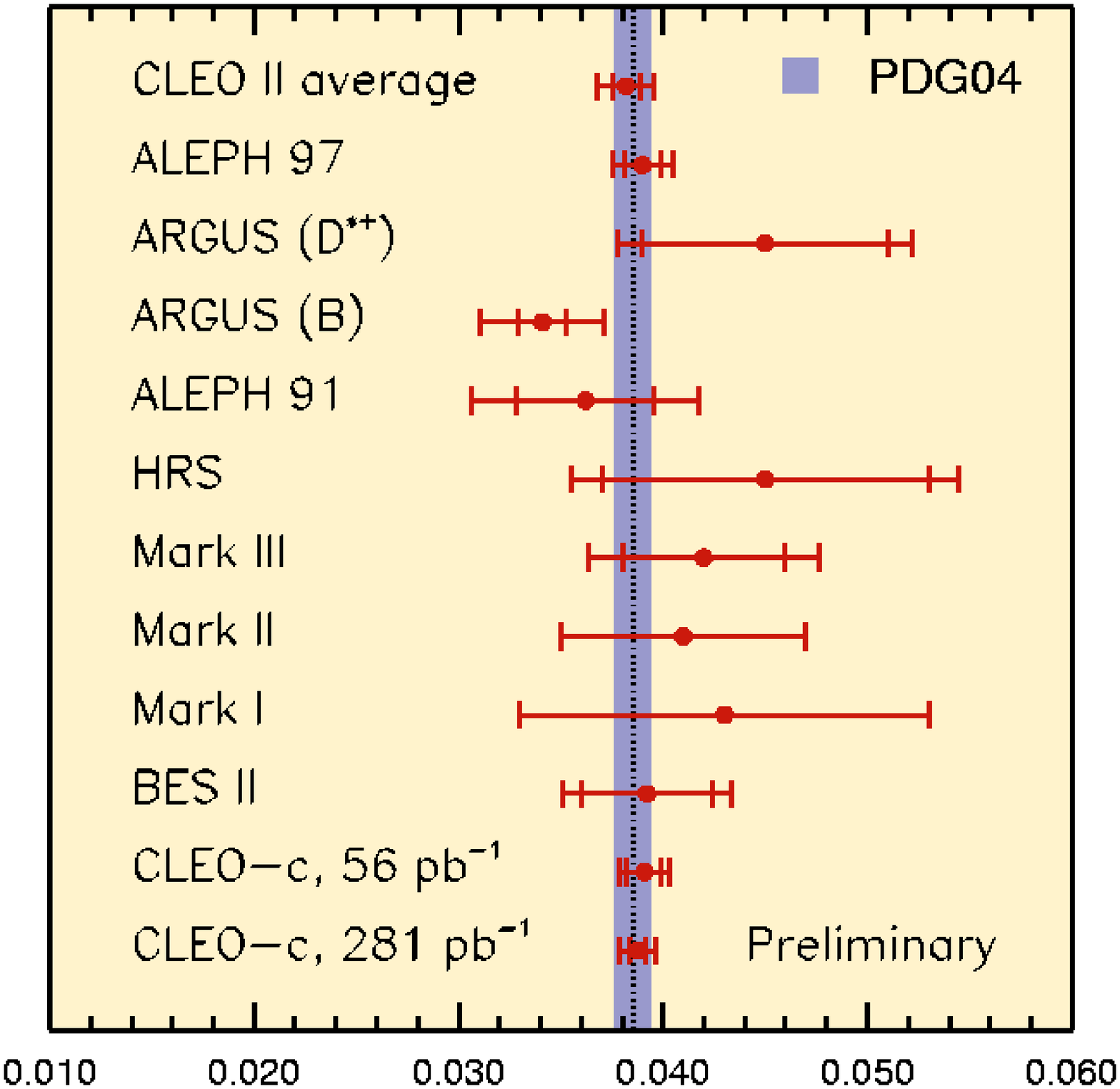}
  \includegraphics[height=2.22in,width=2.5in]{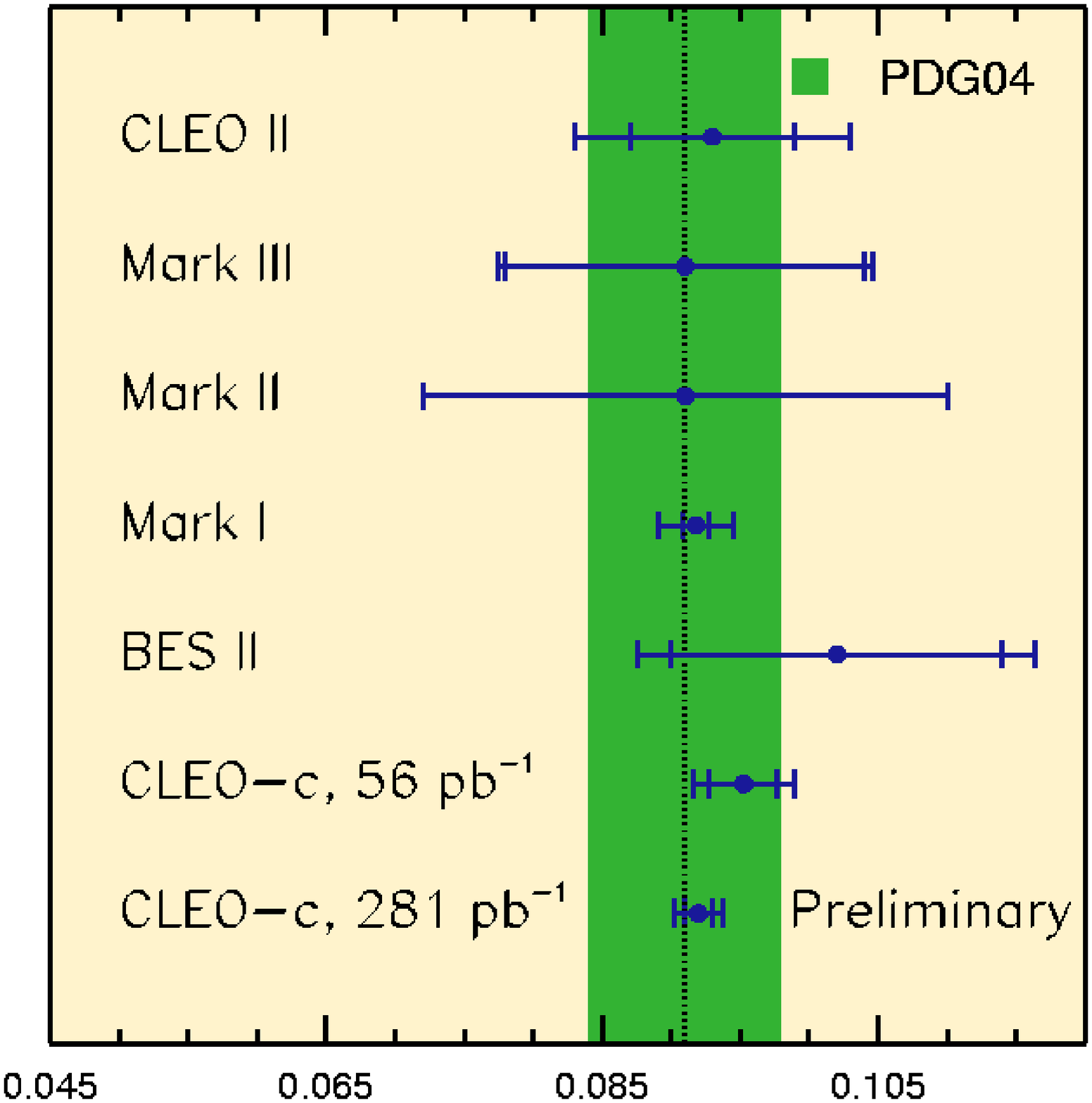}
  \caption{\cleoc absolute \ddokpi (left) and \ddpkpipi (right) branching
           fractions compared with other measurements.}
  \label{DoDpBrComp}
\end{figure}

Two of these branching fractions, Br(\ddokpi) and Br(\ddpkpipi)
shown in Table~\ref{tab:DBranching}, are particularly
important because they have been used to normalized
practically all other \ddo and \ddp modes.
Figure~\ref{DoDpBrComp} shows a comparison of our results
(based on 56~pb$^{-1}$ and 281~pb$^{-1}$),
for these two modes
with other experimental measurements.
These results already
represent significant improvements with respect to the world averages.
To date, they are the most precise measurements of hadronic branching
fractions for D mesons.
Other results based on the latest \psidd data sample are also available,
such as
the semi-leptonic inclusive decays~\cite{Adam:2006ia}.
These high precision results are possible in part by our
detector performance.

\section{T\lowercase{he} D$_S$ E\lowercase{nergy} S\lowercase{can}}\label{ds}
\begin{figure}[!h]
  \centering
  \includegraphics[height=2.2in,width=2.9in]{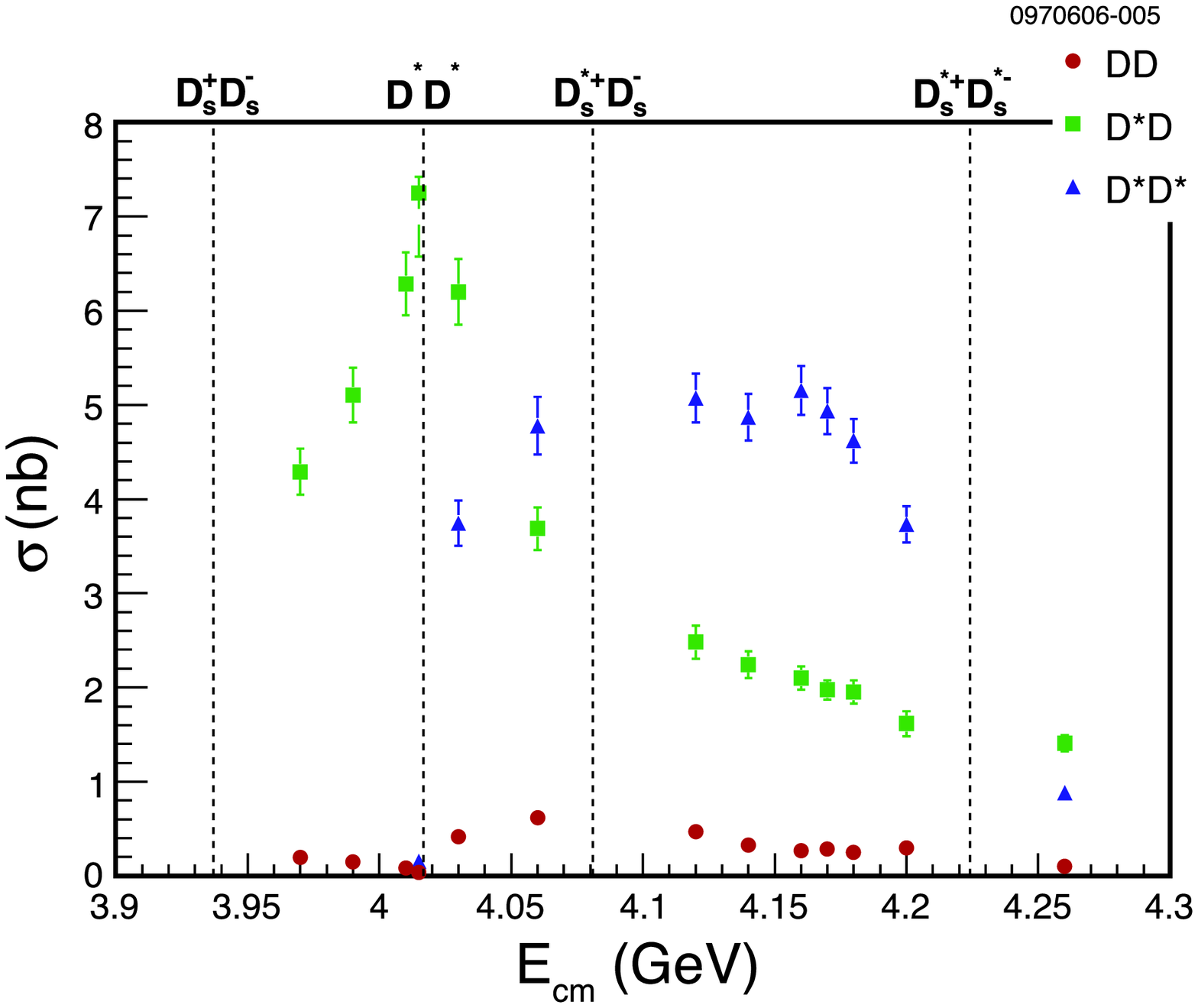}
  \includegraphics[height=2.2in,width=2.9in]{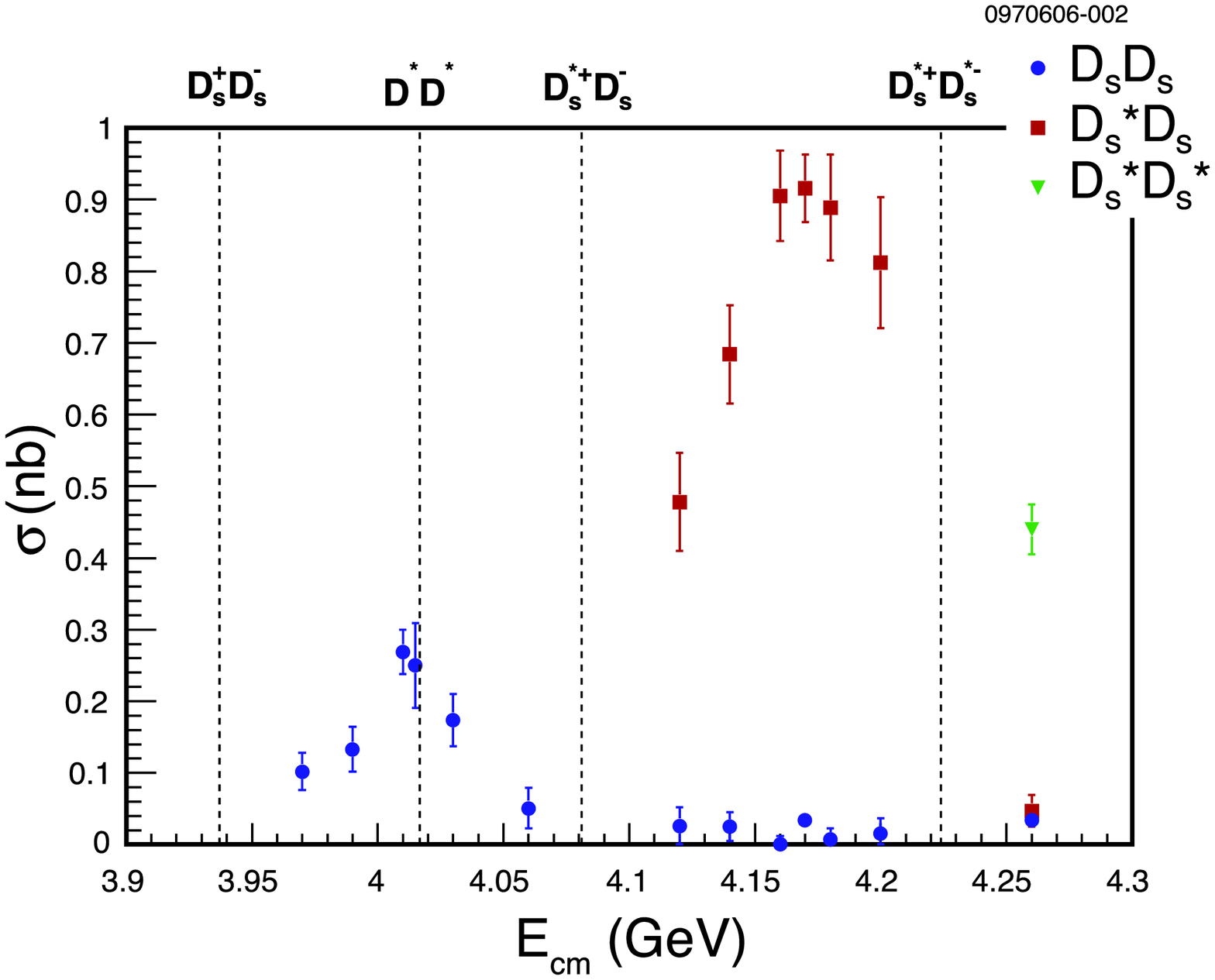}
  \caption{Preliminary measurement production cross section for
         $D\overline{D}$, 
         $D\overline{D^*}$,
         $D^*\overline{D^*}$ on the left and
         $D_s\overline{D_s}$, 
         $D_s\overline{D^*_s}$,
         $D^*_s\overline{D^*_s}$ on right
	 as a function of the \epem center-of-mass energy.}
  \label{scanDD}
\end{figure}

In order to select the optimal energy to study \ddsUns,
a scan on the energy region from 3970-4260~MeV
was performed~\cite{Poling:2006da}.
\begin{figure}[!h]
  \centering
  \includegraphics[height=2.2in,width=4.5in]{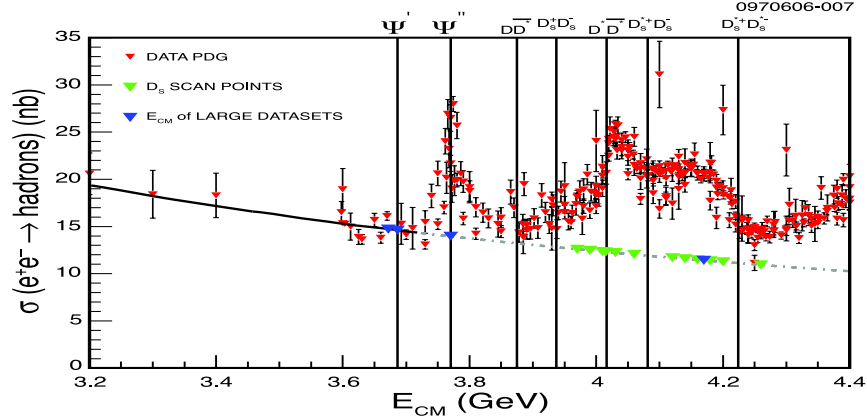}
  \caption{Data on hadron production cross section as
         a function of the \epem center-of-mass energy.
         Inverted green triangles indicates the \cleoc scan run points.}
\label{scanDs}
\end{figure}
The scan consisted on 12 points~(Fig.~\ref{scanDs})
with a total luminosity of 60~pb$^{-1}$.
The last energy point at E$_{cm} = 4260$~MeV,
was taken to study the Y(4260) state (See section~\ref{cc}).
At each energy point the first task was to make quick determination
of the cross section for each of the two charmed meson final states
that were accessible at that energy.
Several methods were used to measure the cross section.
It was measured    
by counting
inclusive hadronic, inclusive D, 
and exclusive DD$^*$ final states events.
Fig.~\ref{scanDD} shows these cross section
using exclusive final states.
A peak cross section of about 1.0~nb for \ddsUns$\overline{\ddsSta}$
is observed around 4170~MeV, which was then selected as the
optimal energy for \cleoc to carry out the
\ddsUns physics program.
At this energy, we produce \ddsP\ddsN pairs, where one of the \ddsUns,
accompanied by a \phot or \pizero,
is in general the daughter of a \ddsSta decay.
In order to avoid a large efficiency loss and
contributions to systematic uncertainties that would
arise from the soft photon,
we make no attempt to find either the \phot or the \pizero.
Thus, the  main variable used in this analysis is the
reconstructed invariant mass of the \ddsUns instead of the m$_{bc}$.

\begin{figure}[!h]
  \centering
  \includegraphics[height=2.20in,width=2.5in]{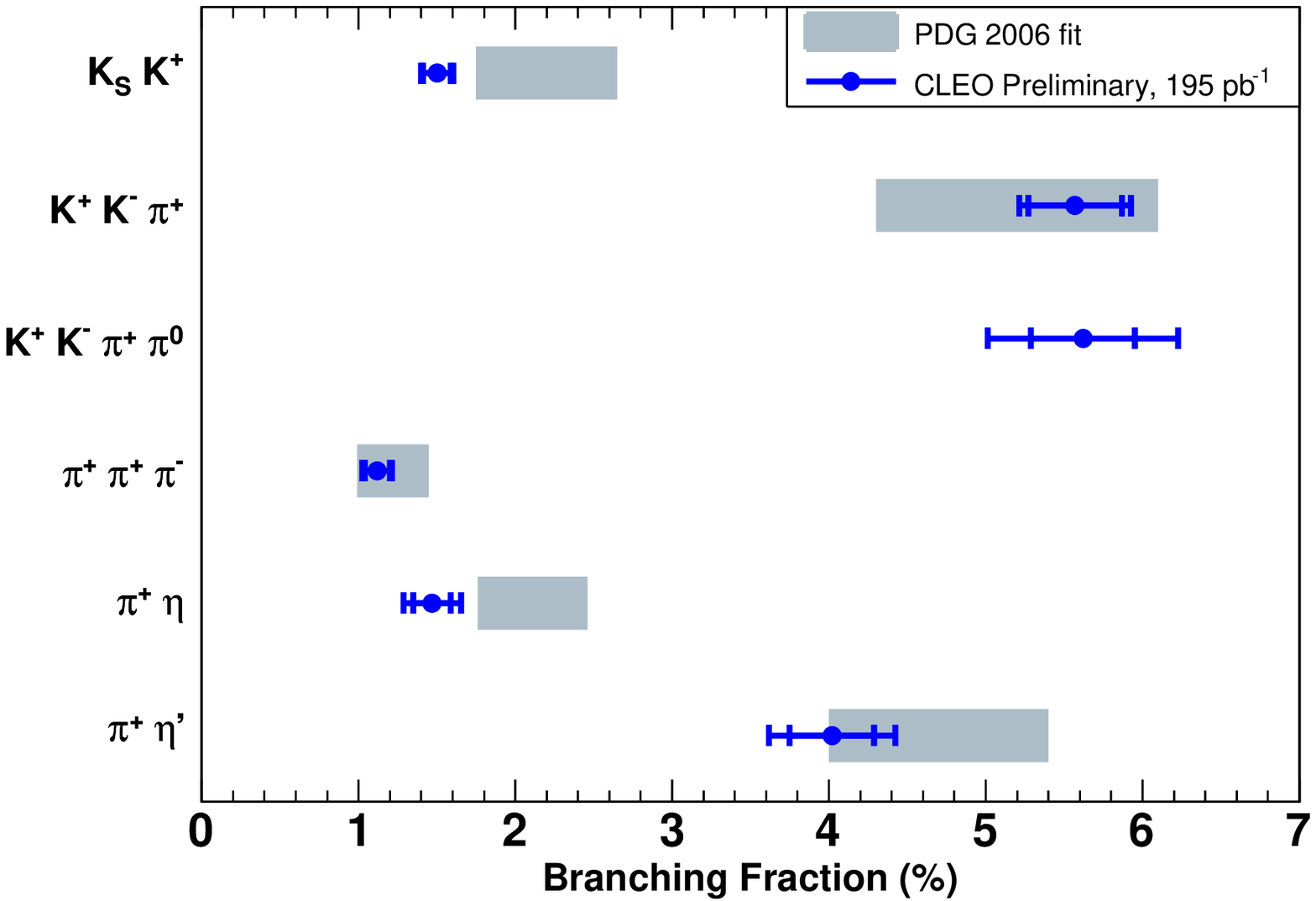}
  \includegraphics[height=2.25in,width=2.5in]{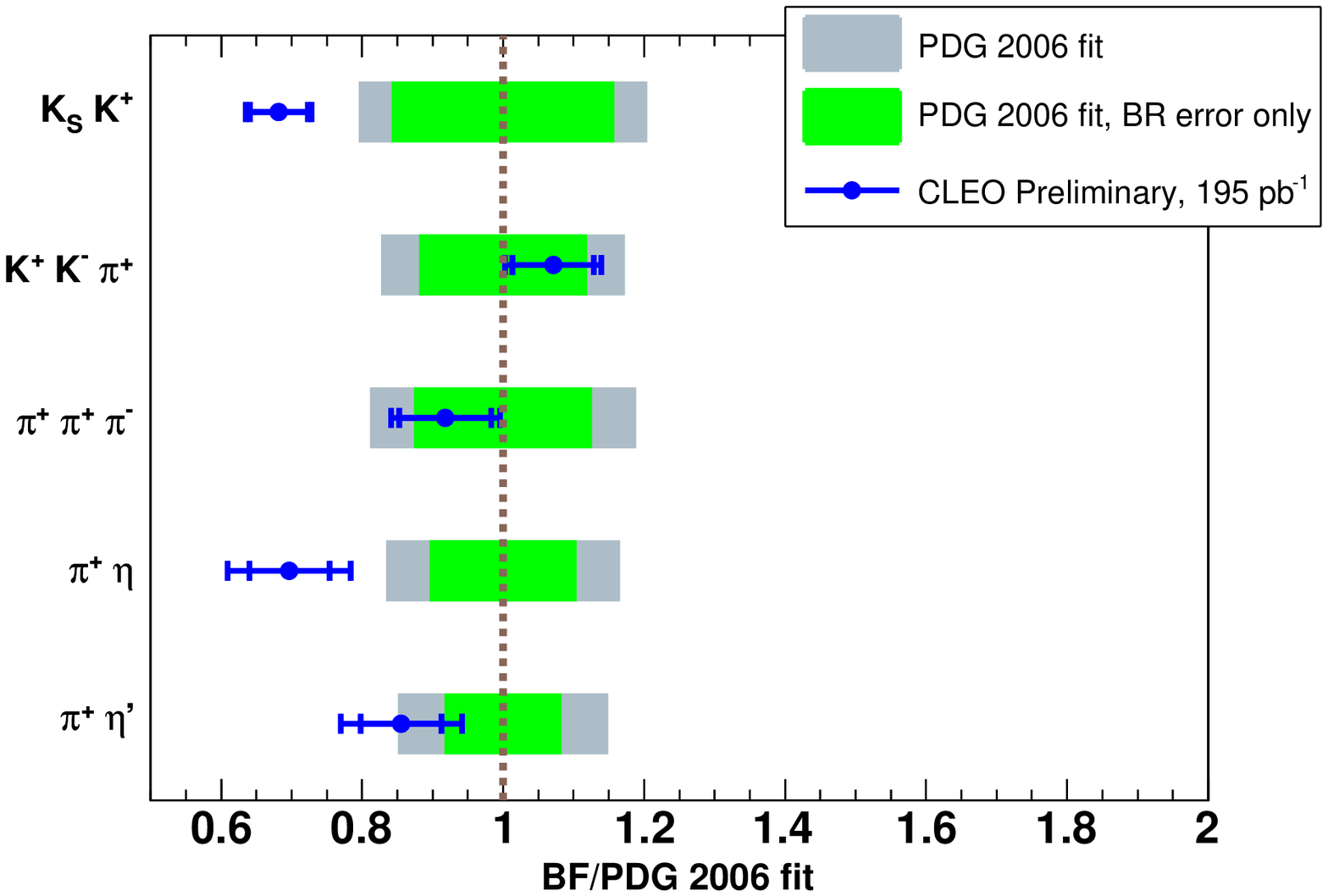}
  \caption{Absolute \ddsP branching ratio on the left and
         our results normalized with respect to the PDG 2006 values on the right.}
  \label{DsBr}
\end{figure}

An initial sample of 
195~pb$^{-1}$ of data
at the center-of-mass energy \comeq~4170~MeV
were recorded.
High precision \ddsUns inclusive decays analysis have been already
published~\cite{Huang:2006nx} and
preliminary measurements on
exclusive absolute hadronic branching fractions have already reported 
for several \ddsP decay modes~(see Table~\ref{tab:DBranching})
by using single \ddsUns tags~\cite{Adam:2006me}.
These measurements are shown in Fig.~\ref{DsBr} and
while these results are preliminary, they
are more precise than the one published on PDG 2006~\cite{PDG2006}
as is shown in Fig.~\ref{DsBr}.

Our Br(\ddsPkkpi) include
the Br(\ddsPphipi) which is one of the largest \ddsUns branching
fractions~\cite{PDG2006}.
It has often used as a reference for other \ddsUns decays.
This measurement has been essentially derived from a
narrow mass region cut around the 
\phip in the \kaka invariant mass (\phidec).
However, 
there is a strong evidence~\cite{Frabetti:1995sg} for a
broad contribution from $f_0(980)$ or $a_0(980)$ under the $\phi\pi^+$ region.
This scalar contribution accounts for approximately 5\%,
which is comparable to our experimental errors for the Br(\ddsPkkpi).
\cleoc will soon exceed this level of precision and the Br(\ddsPphipi)
will be measured
as soon as
the new data is gathered.

\section{C\lowercase{harmonium and} C\lowercase{harmonium}-L\lowercase{ike}
         S\lowercase{tates}}\label{cc}
The spectrum below the \ddmix is well known,
however, in the last few years there has been a
renewed interest in heavy quark spectroscopy.
More than a dozen of unexpected charmonium-like states have been
reported by high luminosity experiments and several long elusive
states have been observed,
including
the \hc~\cite{Rosner:2005ry,Andreotti:2005vu} and
the \etacp~\cite{Asner:2003wv}.  
Many new theoretical models have been proposed to explain
all these resonances.

\begin{figure}     
  \centering
  \includegraphics[height=4.0in,width=4.2in]{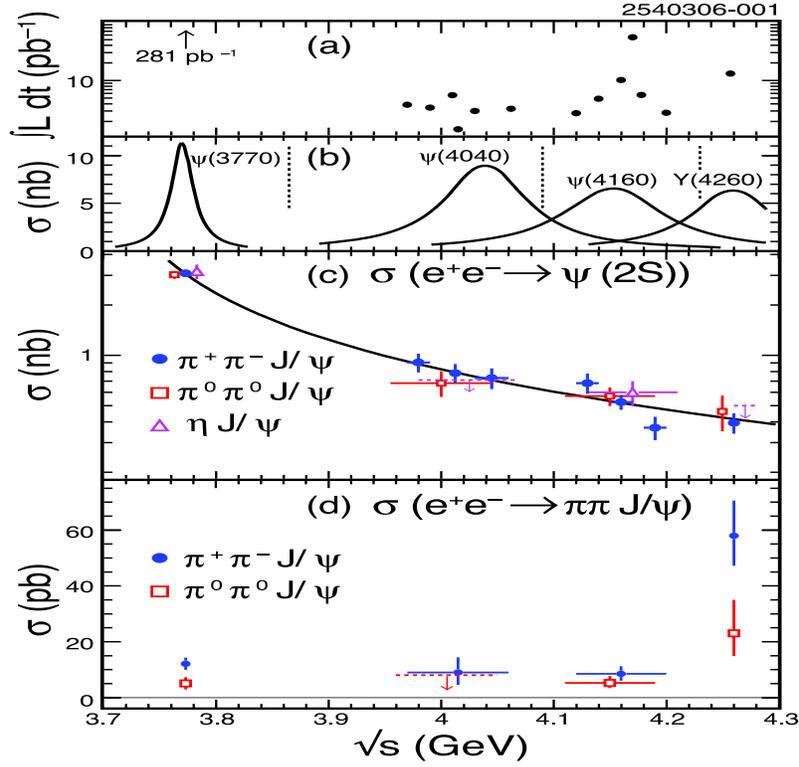}
  \caption{(a) Integrated luminosity vs. $\sqrt{s}$,
           (b) Born-level Breit-Wigner cross section for established charmonium
	       and Y(4260) states.
	   (c) \epem~$\rightarrow$~\phot\psip cross section vs. energy
	       for three different \psip decay mode.
	   (d) \epem~$\rightarrow$~\jpsi$\pi\pi$ vs. $\sqrt{s}$.
	   Some points at (c) and (d) are shifted by 10~MeV for
	   clarity.}
  \label{y4260fig}
\end{figure}

The last energy point of our \ddsUns energy scan, E$_{cm} = 4260$~MeV,
with a total integrated luminosity of 13.2~pb$^{-1}$
was used
primarily to investigate the Y(4260) state discovered by
the BaBar Collaboration~\cite{Aubert:2005rm}.
This observation was based on a sample of 233~fb$^{-1}$
collected at the $\Upsilon(4S)$ in an
initial state radiation events as
\epem~$\rightarrow$~\phot(\jpsi$\pi^+\pi^-$).
\cleoc made the first confirmation~\cite{y4260},
at 11$\sigma$ significance,
of this new charmonium decay mode (\jpsi$\pi^+\pi^-$) of the
Y(4260) state,
made the first observation of
Y(4260)~$\rightarrow$~\jpsi$\pi^0\pi^0$ at 5.1$\sigma$
and
find the first evidence for Y(4260)~$\rightarrow$~\jpsi$K^+K^-$
at 3.7$\sigma$ significance.
The measured \cleoc cross section as well as the integrated luminosity are
shown on Fig.~\ref{y4260fig}.
This signal was also confirmed by using the \cleoiii dataset of 
13.3~fb$^{-1}$ collected at $\Upsilon(1S)-\Upsilon(4S)$ resonances
and
the extracted signal parameters are consistent with
BaBar's results.

The 3686~MeV dataset, taken at the \psip peak included
a sample of approximately
26 million \psip decays acquired during 2006
to explore charmonium spectroscopy
and related states.
In the meantime this sample is being prepared for a detailed analysis,
a small \psip sample of $\sim$5.6~pb$^{-1}$,
corresponding approximately to 3~million \psip decays,
split equally between \cleoiii and \cleoc,
has produced a great  well known variety
of results which included
\jpsi leptonic decays~\cite{Adams:2005mp},
\hc~\cite{Rosner:2005ry} discovery, etc.


The singlet P-state of charmonium \hc has been observed 
in the isospin forbidden reaction
\psip$\rightarrow$\pizero\hc,
where \hc$\rightarrow$\phot\etac.
The \etac was identified by two methods:
first, it was fully reconstructed in 7 exclusive modes,
and second, it was reconstructed inclusively.
The exclusive mode
has the advantage of signal purity while
the inclusive mode has the advantage of larger signal yield.
Both methods yields consistent results and
the average measured mass is $3524.4 \pm 0.6 \pm 0.4$~MeV.
This mass leads to a hyperfine splitting of $1 \pm 0.6 \pm 0.4$~MeV,
which is consistent with zero as expected. 
With this discovery, the charmonium family
below \psidd is complete, and the mass values can be used
in potential models to predict higher states.
A detailed analysis on the \hc observation
made by the experiment
can be found at~~\cite{Rosner:2005ry}.

\section{S\lowercase{ummary}}
Running at charm threshold with D tagging provides
a powerful and "background free" environment
that allow very precise measurements.
Our experimental measurements on leptonic, semi-leptonic and
hadronic D decays are producing a major impact
on many charm decays  with broad implications for all
flavor physics.
These high precision measurements are more precise
than previous world average~\cite{PDG2004}.

The experiment is scheduled to run until March 31 of 2008.
In addition to the charmonium sample, 
we plan to
further increase our samples at \psidd and at 4170~MeV
to continue the open and closed charm measurements.
These high statistics samples from the decays of charmed mesons and
charmonium data represent a rich source of important physics
on the study of charm quarks.
It permits detailed studies of weak and strong interaction physics.







\begin{theacknowledgments}
I would like to thank the organizer for the invitation to
give this talk and for their hospitality during the conference.
I thank my colleagues from the \cleoc Collaboration for the many
useful comments to this paper.
This work was supported in part by the
U.S. Department of Energy.     
\end{theacknowledgments}



\bibliographystyle{aipproc}   

\bibliography{hmendez2_cleo_ref}

\IfFileExists{\jobname.bbl}{}
 {\typeout{}
  \typeout{******************************************}
  \typeout{** Please run "bibtex \jobname" to optain}
  \typeout{** the bibliography and then re-run LaTeX}
  \typeout{** twice to fix the references!}
  \typeout{******************************************}
  \typeout{}
 }

\end{document}

%% file: hmendez2_cleo_def.tex
\def\cleoc{\mbox{CLEO-c}\xspace}
\def\cleoiii{\mbox{CLEO III}\xspace}
\def\comeq{$\sqrt{s} =$}

\newcommand{\phot  }{\ensuremath{\mathbf {\gamma} }\xspace}
\newcommand{\pizero}{\ensuremath{\mathbf {\pi^0} }\xspace}

\newcommand{\etz  }{\ensuremath{\eta}\xspace}
\newcommand{\etp  }{\ensuremath{\etz'}\xspace}
\newcommand{\phip }{\ensuremath{\phi}\xspace}
\newcommand{\ttau }{\ensuremath{\tau}\xspace}

\newcommand{\kshor}{\ensuremath{K^0_S}\xspace}

\newcommand{\psidd }{\ensuremath{\mathbf{\psi}(3770)}\xspace}
\newcommand{\jpsi  }{\ensuremath{J/\mathbf{\psi}}\xspace}
\newcommand{\psip  }{\ensuremath{\mathbf{\psi}(2S)}\xspace}
\newcommand{\hc    }{\ensuremath{h_c(^1P_1)}\xspace}

\newcommand{\etac  }{\ensuremath{\etz_c}\xspace}
\newcommand{\etacp }{\ensuremath{\etz'_c(2^1S_0)}\xspace}

\newcommand{\pipi     }{\ensuremath{\pi^+\pi^-}\xspace}

\newcommand{\kaka     }{\ensuremath{K^+K^-}\xspace}

\newcommand{\kaNpiP   }{\ensuremath{K^-\pi^+}\xspace}
\newcommand{\piodecay }{\ensuremath{\pizero \rightarrow \phot\phot}\xspace}

\newcommand{\phidec   }{\ensuremath{\phip \rightarrow \kaka}\xspace}

\newcommand{\kshordec }{\ensuremath{\kshor\rightarrow \pipi}\xspace}

\newcommand{\etpdec   }{\ensuremath{\etp \rightarrow \pipi\etz}\xspace}

\newcommand{\chicjn}{\ensuremath{\chi_{cJ}~(J=0,1,2)}\xspace}

\newcommand{\ddmix}{\ensuremath{D\overline{D}} \xspace}
\newcommand{\ddo   }{\ensuremath{D^0 }\xspace}

\newcommand{\dds   }{\ensuremath{D^\pm_s}\xspace}
\newcommand{\ddsP  }{\ensuremath{D^+_s}\xspace}
\newcommand{\ddsN  }{\ensuremath{D^-_s}\xspace}
\newcommand{\ddp   }{\ensuremath{D^+}\xspace}
\newcommand{\ddm   }{\ensuremath{D^-}\xspace}
\newcommand{\ddsUns}{\ensuremath{D_s}\xspace}
\newcommand{\ddsSta}{\ensuremath{D^*_s}\xspace}

\newcommand{\epem    }{\ensuremath{e^+e^-}\xspace}
\newcommand{\epemtodd}{\ensuremath{\epem\rightarrow~\psidd\rightarrow~\ddmix}\xspace}
\newcommand{\epemtodz}{\ensuremath{\epem\rightarrow~\psidd\rightarrow~\ddo\overline{\ddo}}\xspace}
\newcommand{\epemtodp}{\ensuremath{\epem\rightarrow~\psidd\rightarrow~\ddp\ddm}\xspace}

\newcommand{\ddokpi    }{\ensuremath{\ddo\rightarrow~\kaNpiP}\xspace}
\newcommand{\ddokpipiz }{\ensuremath{\ddo\rightarrow~\kaNpiP\pizero}\xspace}
\newcommand{\ddokpipipi}{\ensuremath{\ddo\rightarrow~\kaNpiP\pipi}\xspace}

\newcommand{\ddpkpipi   }{\ensuremath{\ddp\rightarrow~\kaNpiP\pi^+}\xspace}
\newcommand{\ddpkpipipiz}{\ensuremath{\ddp\rightarrow~\kaNpiP\pi^+\pizero}\xspace}
\newcommand{\ddpkspi    }{\ensuremath{\ddp\rightarrow~\kshor\pi^+}\xspace}
\newcommand{\ddpkspipiz }{\ensuremath{\ddp\rightarrow~\kshor\pi^+\pizero}\xspace}
\newcommand{\ddpkspipipi}{\ensuremath{\ddp\rightarrow~\kshor\pi^+\pipi}\xspace}
\newcommand{\ddpkkpi    }{\ensuremath{\ddp\rightarrow~\kaka\pi^+}\xspace}

\newcommand{\ddsPksk    }{\ensuremath{\ddsP\rightarrow~\kshor K^+}\xspace}
\newcommand{\ddsPkkpi   }{\ensuremath{\ddsP\rightarrow~\kaka\pi^+}\xspace}
\newcommand{\ddsPkkpipiz}{\ensuremath{\ddsP\rightarrow~\kaka\pi^+\pizero}\xspace}
\newcommand{\ddsPpipipi }{\ensuremath{\ddsP\rightarrow~\pipi\pi^+}\xspace}
\newcommand{\ddsPpietz  }{\ensuremath{\ddsP\rightarrow~\pi^+\etz}\xspace}
\newcommand{\ddsPpietp  }{\ensuremath{\ddsP\rightarrow~\pi^+\etp}\xspace}
\newcommand{\ddsPphipi  }{\ensuremath{\ddsP\rightarrow~\phip\pi^+}\xspace}